\documentclass[aps,twocolumn,pra,showpacs,amsmath,amssymb,showkeys,10pt]{revtex4-2}
\usepackage{graphicx}
\usepackage{epstopdf}
\usepackage{diagbox}
\pdfoutput=1
\usepackage{braket}
\usepackage{hyperref}
\usepackage{longtable}
\newcommand{\tabincell}[2]{\begin{tabular}{@{}#1@{}}#2\end{tabular}}

\begin{document}

	\title{Supercomputer model of finite-dimensional quantum electrodynamics applications}
	
	\author{Wanshun Li}
	\affiliation{Faculty of Computational Mathematics and Cybernetics, Lomonosov Moscow State University, Vorobyovy Gory 1, Moscow, 119991, Russia}
	
	\author{Hui-hui Miao}
	\affiliation{Faculty of Computational Mathematics and Cybernetics, Lomonosov Moscow State University, Vorobyovy Gory 1, Moscow, 119991, Russia}

	\author{Yuri Igorevich Ozhigov}
	\email[Email address: ]{ozhigov@cs.msu.ru}
	\affiliation{Faculty of Computational Mathematics and Cybernetics, Lomonosov Moscow State University, Vorobyovy Gory 1, Moscow, 119991, Russia\\K. A. Valiev Institute of physics and technology, Russian Academy of Sciences, Nakhimovsky Prospekt 36, Moscow, 117218, Russia}

	\date{\today}

	\begin{abstract}
	A general scheme is given for supercomputer simulation of quantum processes, which are described by various modifications of finite-dimensional cavity quantum electrodynamics models, including Jaynes--Cummings--Hubbard model and Tavis--Cummings--Hubbard model. Conclusions and recommendations are illustrated using two examples: approximate model of hydrogen bonding and model of photon motion on a two-dimensional plane.
	\end{abstract}


	\maketitle

	\section{Introduction}
	\label{sec:Intro}
	
	Modeling of quantum processes is one of the most important areas of application of supercomputers. Even calculations of the configurations of complex compounds: proteins and nucleic acids, require the use of high-performance computing; it's particularly true for modeling of dynamics, in which the main difficulty lies in the need to take into account the multi-mode electromagnetic field \cite{Miao2023, Ozhigov2023}. It is known that a computer simulator of chemical reactions has not yet been implemented, due to the prohibitive complexity of the formal description of the states of charges and fields.
	
	To overcome the exponential complexity barrier, R. Feynman proposed building a so-called quantum computer, the operating principle of which is to simulate a real process by the evolution of an artificially created quantum device consisting of quantum bits (qubits) connected in a circuit similar to microelectronic devices \cite{Feynman1982}. Despite the proven mathematical impeccability of quantum computing \cite{Barenco1995, Zalka1998, Wiesner1996}, their physical implementation encounters a fundamental difficulty --- decoherence, that is, the spontaneous decay of quantum states, the degree of which increases with increasing of complexity, so there is an uncertainty relation of the form ''accuracy-complexity''
	\begin{equation}
		\label{eq:A-C}
		A(\Psi)C(\Psi)\leq Q
	\end{equation}
	the constant $Q$ in which, equals to the maximum number of qubits of a quantum processor, judging by experiments, doesn't exceed several decades \cite{Ozhigov2022}. Thus, the modeling of complex processes at the predictive (quantum) level turns out to be, in principle, within the capabilities of existing supercomputers, and the only task is to build mathematical software for this.
	
	The specificity of such tasks requires their allocation to a special class, which is necessary for planning the distribution of time and memory for the operation of a multi-user supercomputer complex, as well as the necessary software libraries. In recent years, some supercomputer simulation studies of quantum dynamics \cite{Zhao2021, Shang2022} have been carried out. In this paper, we will give some examples of supercomputer solutions of such problems based on standard quantum electrodynamics (QED) models and give recommendations for scaling them to even more complex processes at the boundaries of biology.
	
	This paper is organized as follows. After introducing the general scheme for computer modeling of quantum processes in Sec. \ref{sec:General_scheme}, we introduce algorithms and their features in Sec. \ref{sec:Algorithms}. We investigate the approximate model of hydrogen bonding in Sec. \ref{subsec:ApproxHB} and the model of photon motion on a two-dimensional plane in Sec. \ref{subsec:PhotonMotion2D}. Some brief comments on our results in Sec. \ref{sec:Conclusion} close out the paper.
	
	\section{General scheme for computer modeling of quantum processes}
	\label{sec:General_scheme}
	
	\begin{figure*}
		\centering
		\includegraphics[width=0.6\textwidth]{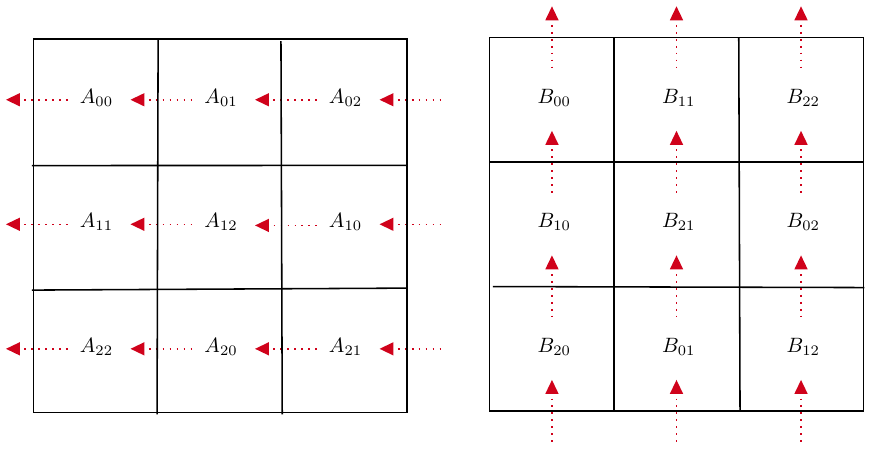} 
		\caption{(online color) Transferring operation of Cannon's algorithm after completing the alignmentation.}
		\label{fig:cannon_algroithm}
	\end{figure*}
	
	The general scheme for modeling of complex processes at the quantum level consists of four points
	\begin{itemize}
	\item Establishing the general structure of the basic quantum states of the entire system "matter $+$ field" as a series of qubits, fixing their physical interpretation, and creating the matrix of the system's energy operator $H$ (Hamiltonian).
	\item Determination of the essential physical factors of decoherence $A_j$ that determine the irreversible interaction of the system under consideration with the environment and writing a specific form of the quantum master equation (QME)
	\begin{equation}
		\label{eq:QME}
		i\hbar\dot{\rho}=[H,\rho ]+iL(\rho)
	\end{equation}
	where $\rho$ is the density matrix of the system being studied and
	\begin{equation}
		\label{eq:QME1}
		L(\rho)=\sum\limits_j\gamma_j\left(A_j\rho A_j^\dagger-\frac{1}{ 2}\left\{ A_j^\dagger A_j\right\}\right)
	\end{equation}
	where $j$ takes no more than three values, but in the general case the system of operators $A_j$ supplemented by a unit operator constitutes an orthonormal basis of the Liouville space of operators with the scalar product $\langle A|B\rangle=tr(A^\dag B)$, where $\dag$ denotes Hermitian conjugate.
	\item Approximate solution of the equation \eqref{eq:QME} using the method of selecting of working space (subspace containing the approximate solution).
	\item Finding trial values of the Hamiltonian parameters using general multi-criteria optimization methods (neural networks, evolutionary genetic algorithms) to match the experimental results.
	\end{itemize}
	
	In this work, we will touch only upon the first three points of the program --- using the example of the application of modified QED models to chemical transformations.
	
	\section{Algorithms and their features}
	\label{sec:Algorithms}
	
	\subsection{Cannon's algorithm}
	\label{subsec:CannonAlgorithm}
	
	\begin{figure*}
		\centering
		\includegraphics[width=0.6\textwidth]{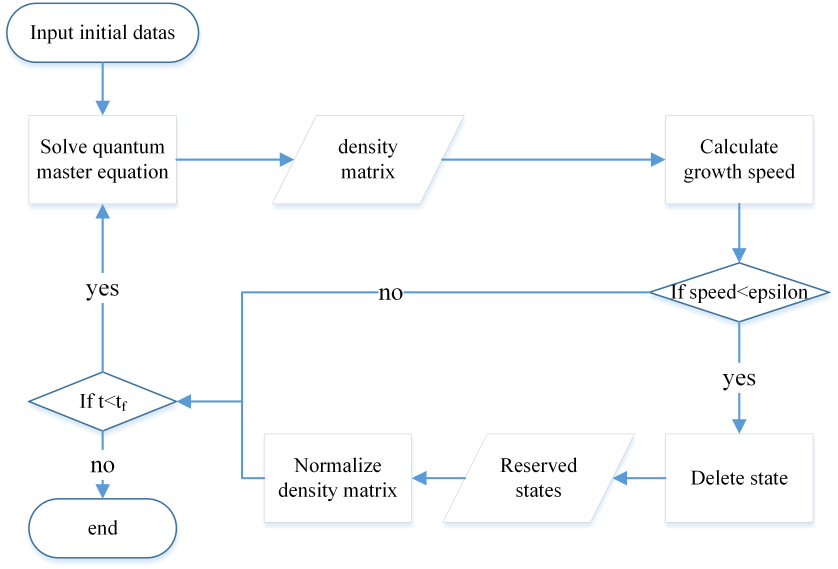} 
		\caption{(online color) Scheme of state space selection algorithm.}
		\label{fig:selection_algroithm}
	\end{figure*}
	
	In computer science, Cannon's algorithm is a distributed algorithm for matrix multiplication for two-dimensional meshes first described by L.E. Cannon \cite{Cannon1969}. In order to better understand the application of the Cannon's algorithm in this paper, we take the approximate model of hydrogen bonding, described in Sec. \ref{subsec:ApproxHB}, as an example. The basic state of the model with two hydrogen bonds occupies 12 qubits and generates $2^{12}$ different states. Therefore, the dimension of all operators also equals to $2^{12}$. According to the Hamiltonian described in Eq. \eqref{eq:hamiltonian}, in our model there are 7 different operators, and when solving the quantum master equation, the number of iterations is 3000. In each iteration, it is necessary to perform intensive computational operations with matrices, which include a large number of floating-point arithmetic and memory accesses. Supercomputers typically equipped with large-capacity memory and high-speed network connections and can process large numbers of computing tasks in parallel. For large-scale matrix operations, exploiting the parallel computing capabilities of supercomputers can greatly improve computation efficiency and reduce computation time.
	
	For parallel processing of large matrices, Cannon's algorithm is particularly suitable due to its ability to efficiently process large matrices. It works by dividing matrices into square blocks, distributing those blocks across different processors in a parallel computing environment, and then performing the multiplication operation locally. This approach significantly reduces memory requirements and computation time, making solving QME possible for more complex systems.
	
	Assume that dimension of matrix is $N$, the number of processes is $p$. In the classical Cannon's algorithm, matrices $A$ and $B$ are both divided into $p$ blocks with dimension $N/\sqrt{p}$. Each process is loaded with a blocks $A_{ij}$ and $B_{ij}$, where $0\le i,j<\sqrt{p}$. The calculation scheme is shown as follows
	
	\begin{itemize}
		\item Alignmentation: shift block $A_{ij}$ to the left $i$ steps, shift block $B_{ij}$ up $j$ steps, then process performs multiplication-plus operation;
		\item Transferring shown in Fig. \ref{fig:cannon_algroithm}: shift block from matrix $A$ one step to the left and get the another block from his right neighbour cyclically, shift block from matrix $B$ up one step and get the another block from the neighbour below him cyclically, then process performs multiplication-plus operation;
		\item Repeat "transferring" operation $\sqrt{p}$ times.
	\end{itemize}
	
	\subsection{State space selection algorithm}
	\label{subsec:Space_selection}
	
	Memory on a supercomputer is huge, but it's not unlimited. With an increase in the number of particles in a quantum system, the dimension of the matrix will increase exponentially. We propose an state space selection algorithm \cite{Chen2023} that operates on the probability distribution. The algorithm diagram is shown in Fig. \ref{fig:selection_algroithm}
	
	\begin{itemize}
		\item Construction: construct Hamiltonian, initial density matrix $\rho_0$ and operators for the QME, etc.;
		\item Solving: solve the QME and get the new density matrix;
		\item Calculation: calculate the growth speed of the probability of each state;
		\item Comparison: define in advance the value $\varepsilon$; comparison the growth speed of each state with $\varepsilon$ and if the growth speed of a state is less than $\varepsilon$, this state will be deleted; finally the density matrix will be normalized and the remaining states will be sent to the next iteration;
		\item Repeat the above three operations "solving", "calculation" and "comparison";
		\item Exit: when the iteration goes up to $t_f$, end loop and we get a reduced matrix of key states (selected states).
	\end{itemize}
	
	For the approximate model of hydrogen bonding, the key parameters are presented in the following Tab. \ref{tab:result_of_algorithm}. During the calculation, the algorithm can dynamically remove those states that don't participate in the dynamics actively. This allows us to effectively reduce memory consumption and get approximate results. As shown in the table, the algorithm significantly reduces the dimension of the matrix, which makes it possible to transfer calculations that could only be performed on a supercomputer to a laptop. Then transferring this algorithm to a supercomputer will make it possible to solve problems that would be inaccessible with a standard approach.
	
	\begin{table}
		\centering
		\begin{tabular}{|c|c|c|}
			\hline
			\diagbox{$\varepsilon$}{new size}{original size}& 64$\times$64 & 4096$\times$4096 \\
			\hline
			1e-35 & 45$\times$45 & 272$\times$272 \\
			\hline
			1e-20 & 22$\times$22 & 134$\times$134 \\  
			\hline
		\end{tabular}
		\caption{Effect comparison of state space selection algorithm.}
		\label{tab:result_of_algorithm}
	\end{table}
	
	\section{Finite-dimensional QED applications}
	\label{sec:Applications}
	
	\subsection{Approximate model of hydrogen bonding}
	\label{subsec:ApproxHB}
	
	\begin{figure*}
		\centering
		\includegraphics[width=1.\textwidth]{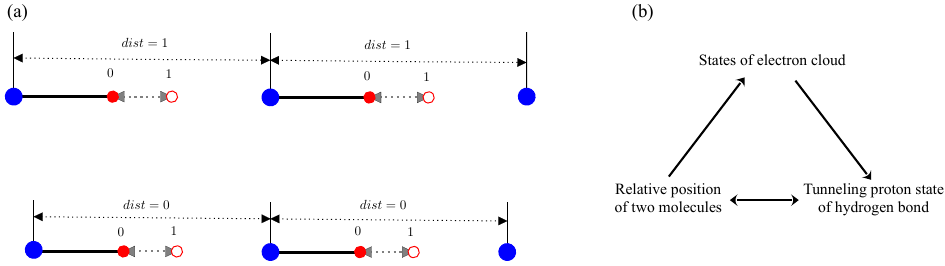} 
		\caption{(online color) Scheme of hydrogen bond formation shown in (a), where the blue dots represent oxygen atoms, the red dots represent hydrogen atoms, and the solid lines between them represent covalent bonds; mutual influence of parts of the model is shown in (b).}
		\label{fig:cova_interpretation}
	\end{figure*}
	
	When water molecules approach each other, an interaction occurs between them and a hydrogen bond will formed. We study a chain of water molecules, the diagram is shown in Fig. \ref{fig:cova_interpretation} (a). As shown in the figure, inside a molecule a hydrogen atom (actually a proton) can tunnel between positions with absorption or emission of a phonon with the mode $\omega_{tun}$, while maintaining a covalent bond. The state of the tunneling proton close to the parent molecule is denoted by $|0\rangle_{proton}$, and the state of the tunneling proton far away from it is denoted by $|1\rangle_{proton}$. The Hamiltonian of this process in the basis $|0\rangle_{proton}$ and $|1\rangle_{proton}$ has the form
	\begin{equation}
		\label{eq:tun-hyd}
		\tilde H_{tun}=\begin{array}{c@{\hspace{-5pt}}l}
			 \begin{array}{c}
			 	|0\rangle_{proton} \\
			 	|1\rangle_{proton} \\
			 \end{array}
			 & \left(
			 \begin{array}{cc}
			 	c_0 & a \\
			 	\bar{a} & c_1 \\
			 \end{array}
			 \right)
			\end{array}
	\end{equation}
In what follows, we consider the model to be symmetrical: $a<0,\ c_0<c_1$. We denote the distance between molecules by $|d\rangle_{dist}$, $d=1$ --- molecules are far away, $d=0$ --- are close. A hydrogen bond is formed only when the molecules are close together. Let's move on to the basis of the eigenstates of this Hamiltonian $|\Phi_0\rangle_{proton}$, $|\Phi_1\rangle_{proton}$, in which it is convenient to write the Jaynes--Cummings scheme \cite{Jaynes1963} with the absorption or emission of a phonon with the mode $\omega_{hyd}$. They have the following form
	 \begin{subequations}
		\label{eq:hybridization}
		\begin{align}
			|\Phi_0\rangle_{proton} & = \frac{1}{\sqrt{\alpha^2+\beta^2}}\left(\alpha|0\rangle_{proton}+\beta|1\rangle_{proton}\right)\label{eq:hybridization0}\\
			|\Phi_1\rangle_{proton} & = \frac{1}{\sqrt{\alpha^2+\beta^2}}\left(-\beta|0\rangle_{proton}+\alpha|1\rangle_{proton}\right)\label{eq:hybridization1}
		\end{align}
	\end{subequations}
	where $\alpha=\beta$. The formation of a hydrogen bond occurs only in the $|\Phi_0\rangle$ state.
	
	Conditional electrons also play important roles in the formation of hydrogen bonds. Electrons have a ground and excited state $|0\rangle_{elec}$ and $|1\rangle_{elec}$, a transition between states with the absorption or emission of a photon with the mode $\omega_e$. A proton can tunnel by interacting with its phonon only if the electron is in its ground state. We will get a diagram of the mutual influence of the parts of the model as shown in Fig. \ref{fig:cova_interpretation} (b): the movement of molecules occurs when the electron is in an excited state and there is no hydrogen bond; proton tunneling between molecules occurs when the molecules are close; electron state transition occurs when a hydrogen bond exists.
	
	The basic states have the following form
	\begin{equation}
		\label{eq:basis_states}
		|m\rangle_{\omega_{tun}}|l\rangle_{proton}|n\rangle_{\omega_{hyd}}|p\rangle_{\omega_e}|e\rangle_{elec}|d\rangle_{dist}
	\end{equation}
	where $m$ --- number of phonons with mode $\omega_{tun}$ interacting with a proton when the molecules are far away ($d=1$); $l$ --- proton state; $n$ --- number of phonons with mode $\omega_{hyd}$ interacting with a proton when the molecules are close ($d=0$); $p$ --- number of photons interacting with electron; $e$ --- electron state; $d$ --- relative location of molecules, $d=0$ --- close, $d=1$ --- far away. Hamiltonian has the following form
	\begin{equation}
		\label{eq:hamiltonian}
		H_{HB}=H_{dist}+H_{elec}+H_{tun}+H_{hyd}
	\end{equation}
	where
	\begin{subequations}
		\label{eq:hamiltonians}
		\begin{align}
			H_{dist} &= \hbar\omega_d\sigma_d^\dagger\sigma_d+g_d\left(\sigma_d+\sigma_d^\dagger\right)\label{eq:hamiltonian_dist}\\
			H_{elec} &= \hbar\omega_e\left(a^\dagger a + \sigma_e^\dagger\sigma_e\right)+g_e\left(a\sigma_e^\dagger+a^\dagger\sigma_e\right)\label{eq:hamiltonian_e}\\
			H_{tun} &= \hbar\omega_{tun}\left(b_{tun}^\dagger b_{tun}+\sigma_{tun}^\dagger\sigma_{tun}\right)\nonumber\\
			&+g_{tun}\left(b_{tun}\sigma_{tun}^\dagger+b_{tun}^\dagger \sigma_{tun}\right)\label{eq:hamiltonian_tun}\\
			H_{hyd} &= \hbar\omega_{hyd}\left(b_{hyd}^\dagger b_{hyd}+ \sigma_{hyd}^\dagger\sigma_{hyd}\right)\nonumber\\
			&+g_{hyd}\left(b_{hyd}\sigma_{hyd}^\dagger+b_{hyd}^\dagger\sigma_{hyd}\right)\label{eq:hamiltonian_hyd}
		\end{align}
	\end{subequations}
	where $a,\ a^\dagger$ --- creation and annihilation operators of photon; $b,\ b^\dagger$ --- creation and annihilation operators of phonon. In this work we simulate a model with two molecules and another model with three. For the three-molecule model, there are two hydrogen bonds, according to Eq. \eqref{eq:basis_states}, that is, the basic state has following form
	\begin{equation}
		\label{eq:basis_states_3mol}
		\begin{aligned}
		&\left(|m_1\rangle_{\omega_{tun}}|l_1\rangle_{proton}|n_1\rangle_{\omega_{hyd}}|p_1\rangle_{\omega_e}|e_1\rangle_{elec}|d_1\rangle_{dist}\right)_1\otimes\\
		&\left(|m_2\rangle_{\omega_{tun}}|l_2\rangle_{proton}|n_2\rangle_{\omega_{hyd}}|p_2\rangle_{\omega_e}|e_2\rangle_{elec}|d_2\rangle_{dist}\right)_2
		\end{aligned}
	\end{equation}
	Hamiltonian is constructed similarly to Eqs. \eqref{eq:hamiltonian} and \eqref{eq:hamiltonians}.
	
	\begin{figure*}
		\centering
		\includegraphics[width=0.9\textwidth]{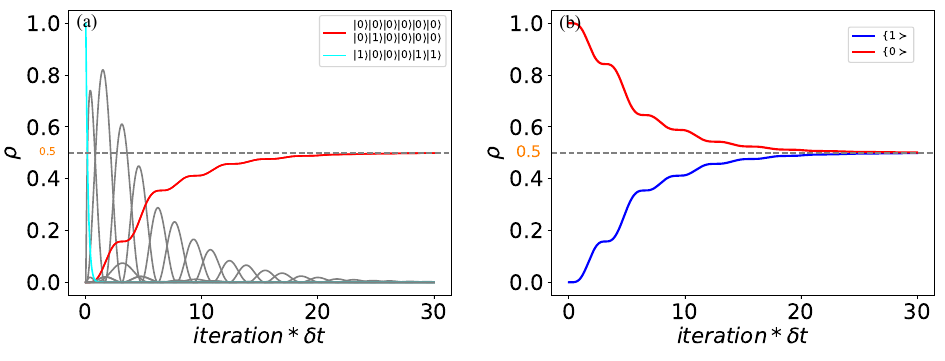} 
		\caption{(online color) Result with one hydrogen bond. The left picture shows the probability dynamics of all states, the cyan line represents the initial state, the red line represents the final state, other states are indicated by gray lines; the right picture shows the probability dynamics in a simplified form: $\{0\succ,\ \{1\succ$.}
		\label{fig:result_one_cova}
	\end{figure*}
	
	\begin{figure*}
		\centering
		\includegraphics[width=0.9\textwidth]{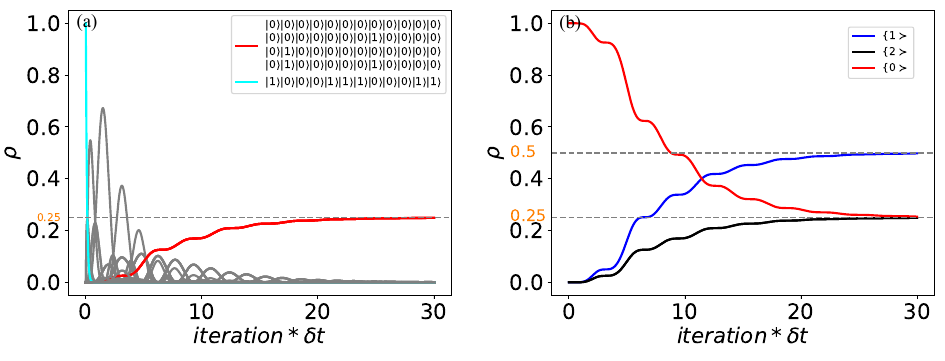} 
		\caption{(online color) Result with two hydrogen bonds. The left picture shows the probability dynamics of all states; the right picture shows the probability dynamics in a simplified form: $\{0\succ,\ \{1\succ,\ \{2\succ$.}
		\label{fig:result_two_cova}
	\end{figure*}
	
	We suppose $\{0\succ$ --- absence of hydrogen bond, $\{1\succ$ --- existence of one hydrogen bond and $\{2\succ$ --- existence of two hydrogen bonds. Fig.\ref{fig:result_one_cova} shows results with one hydrogen bond, Fig.\ref{fig:result_two_cova} shows results with two hydrogen bonds. As shown in these figures, at $iteration*\delta t\to\infty$, in the model with two molecules, the probabilities of $\{0\succ$ and $\{1\succ$ both tends to $\frac{1}{2}$. In a model with three molecules, the probability of $\{1\succ$ tends to $\frac{1}{2}$, the probability of $\{2\succ$ tends to $\frac{1}{4}$, and there is also a probability of $\frac{1}{4}$ that there will be no hydrogen bond.
	
	\begin{table}
		\centering
		\begin{tabular}{|c|c|c|c|}
			\hline
			\tabincell{c}{number\\of nodes} & \tabincell{c}{number of processes\\per node} & \tabincell{c}{total\\processes} & \multicolumn{1}{c|}{run time} \\ \hline
			10                                    & 14                                              & 140                                & 48min14.0426sec               \\ \hline
			20                                    & 14                                              & 280                                & 25min58.5984sec               \\ \hline
			40                                    & 7                                               & 280                                & 27min20.1228sec               \\ \hline
			40                                    & 14                                              & 560                                & 19min30.4321sec               \\ \hline
		\end{tabular}
		\caption{Efficiency comparison table.}
		\label{tab:runtime_table}
	\end{table}
	
	To apply distributed data processing, the matrices are divided into small blocks. The speedup of calculation depends on the number of processes. The Tab. \ref{tab:runtime_table} shows the dependence of the calculation time on the allocation of processes. As shown in the table, when comparing columns 1 and 2, it is clear that with an increase in the number of processes, the cost time is significantly reduced. When comparing columns 2 and 4, with the increase of number of processes, the cost time decreases, but not significantly. This is due to the fact that communication between nodes takes a long time. And comparing columns 2 and 3, which have the same number of processes but different numbers of nodes, we can notice a slight difference in cost time. Communication within nodes is naturally faster than communication between nodes due to shared memory.

	\subsection{Photon motion on a two-dimensional plane}
	\label{subsec:PhotonMotion2D}
	
	\begin{figure*}
		\centering
		\includegraphics[width=0.8\textwidth]{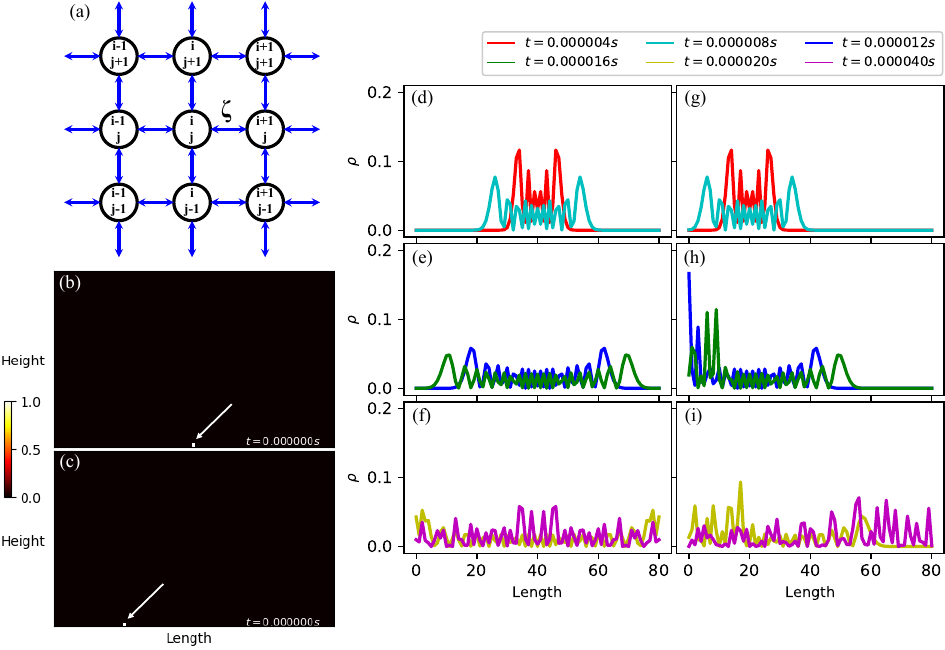} 
		\caption{(online color) The model of photon motion on a two-dimensional plane is shown in (a). (b) and (c) show two cases of initial states: symmetric and asymmetric. (d) $\sim$ (f) show the probability distribution along the length axis in the case of a symmetric initial state. (g) $\sim$ (i) show the probability distribution along the length axis in the case of an asymmetric initial state. And their probability distributions on a two-dimensional plane are shown in Figs. \ref{fig:PhotonMotionSymm} and \ref{fig:PhotonMotionAsymm}, respectively.}
		\label{fig:ModelPhotonMotion2D}
	\end{figure*}
	
	In this section we assume a modified version of the Jaynes--Cummings--Hubbard model (JCHM) \cite{Jaynes1963, Angelakis2007}, in which the photon interacts with a large number of optical cavities that form a two-dimensional plane. The plane of the optical cavities is shown in Fig. \ref{fig:ModelPhotonMotion2D} (a), where the cavity is connected to its neighbors only through optical fibers. There is only one free photon in the system, and it can tunnel from one cavity to another with a tunneling force $\zeta$. Photon leakage is not considered, so the system is closed.
	
	The basic state is shown as follows
	\begin{equation}
		\label{eq:BasicStates}
		|\Psi\rangle=\bigotimes_{i,j}|p_{i,j}\rangle_{i,j}
	\end{equation}
	where $i$ and $j$ represent the coordination of the cavity along the length and height axes, respectively, and $0\leq i\leq\text{Length},\ 0\leq j\leq\text{Height}$, where Length --- number of optical cavities in the horizontal direction (length of the plane), Height --- number of optical cavities in the vertical direction (height of the plane). $p_{i,j}=0$ --- there is no photon inside the cavity, $p_{i,j}=1$ --- there is a photon inside the cavity. The dimension of Hilbert space is $2^{\text{Area}}$, where $\text{Area}=\text{Length}\times \text{Height}$ is the number of optical cavities in a two-dimensional plane (area of plane). The basic state can be defined in another form
	\begin{equation}
		\label{eq:BasicStatesSimple}
		|\Psi\rangle=|i\rangle|j\rangle
	\end{equation}
	where $i$ and $j$ represent the coordination of the cavity within which the photon exists. Thus, the dimension of a Hilbert space is simply the product of the length and the height.
	
	The Hamiltonian of the system is represented by the energy operator in the case of the rotating wave approximation ($\zeta\ll \hbar\omega$)
	\begin{equation}
		\label{eq:Hamiltonian}
		\begin{aligned}
		H_{JCHM}^{RWA}&=\sum_{i,j}^{\text{Length},\text{Height}}\hbar\omega a_{i,j}^{\dag}a_{i,j}\\
		&+\zeta\sum_{\substack{i_1>i_2,j_1>j_2\\i_1-i_2+j_1-j_2=1}}^{\text{Length},\text{Height}}\left(a_{i_1,j_1}^{\dag}a_{i_2,j_2}+a_{i_1,j_1}a_{i_2,j_2}^{\dag}\right)
		\end{aligned}
	\end{equation}
	where $\hbar$ is the reduced Planck constant, $\omega$ is the photonic mode, $a^{\dag}$ is the photon creation operator, $a$ is the photon annihilation operator, and $\zeta$ --- tunneling force.
	
	Now let us assume two cases of initial states: symmetric and asymmetric. In our model, the length of the plane = 81 and the height of the plane = 51, so the area is 4131 and the positions of the four corners are $(0,0)$, $(80,0)$, $(0,50)$ and $( $80.50) The symmetric initial state is at $(40,0)$, as shown in Fig. \ref{fig:ModelPhotonMotion2D} (b). This means, firstly, the photon is in a cavity with coordination: $i=40,\ j=0$. And the asymmetric initial state is in $(20,0)$, as shown in Fig. \ref{fig:ModelPhotonMotion2D} (c).
	
The unitary evolution of the symmetric case is shown in Fig. \ref{fig:ModelPhotonMotion2D} (d) $\sim$ (f). In the case of a symmetric condition, the probability distribution along the length axis is always symmetric. The probability of the appearance of photon gradually spreads from the center in both directions. As the diffusion range increases, the maximum value (peak) of the probability curve tends to decrease. This shows that as time increases, the photon tends to appear uniformly in each cavity. In particular, when the boundary is reached, the probability curve rebounds. In the case of the asymmetric condition, we find that the probability curve first reaches the left boundary and then the right boundary. This causes the probability distribution to become asymmetrical and irregular. Such case is random and unpredictable, which allows photon to serve as carrier of quantum information.

	\begin{figure}
		\centering
		\includegraphics[width=0.5\textwidth]{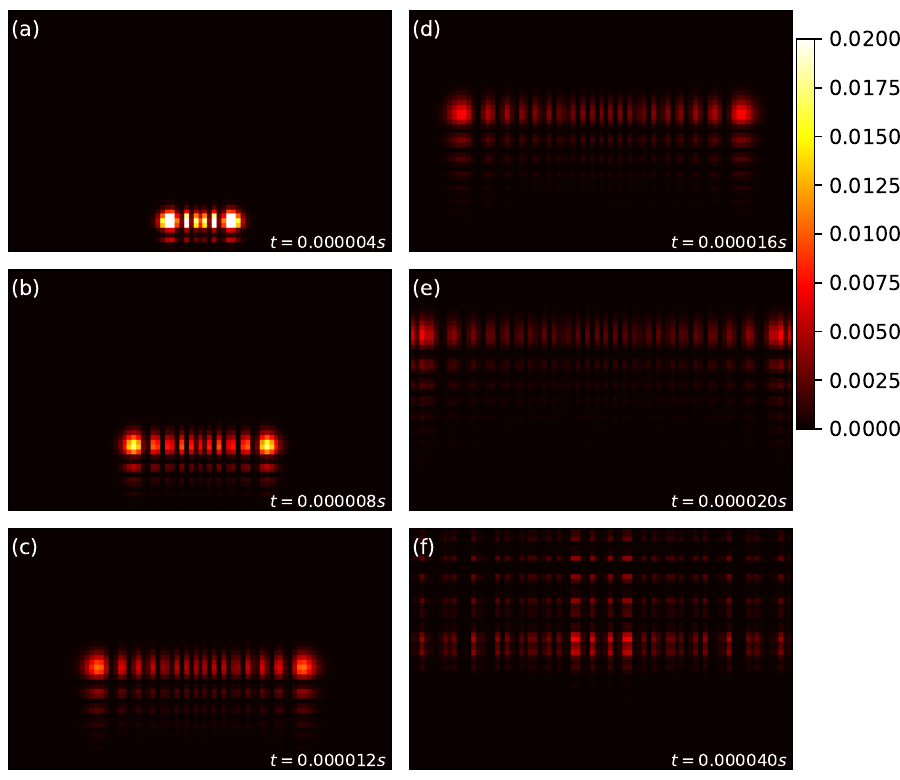} 
		\caption{(online color) Probability distributions in the case of a symmetric initial state.}
		\label{fig:PhotonMotionSymm}
	\end{figure}

\begin{figure}
		\centering
		\includegraphics[width=0.5\textwidth]{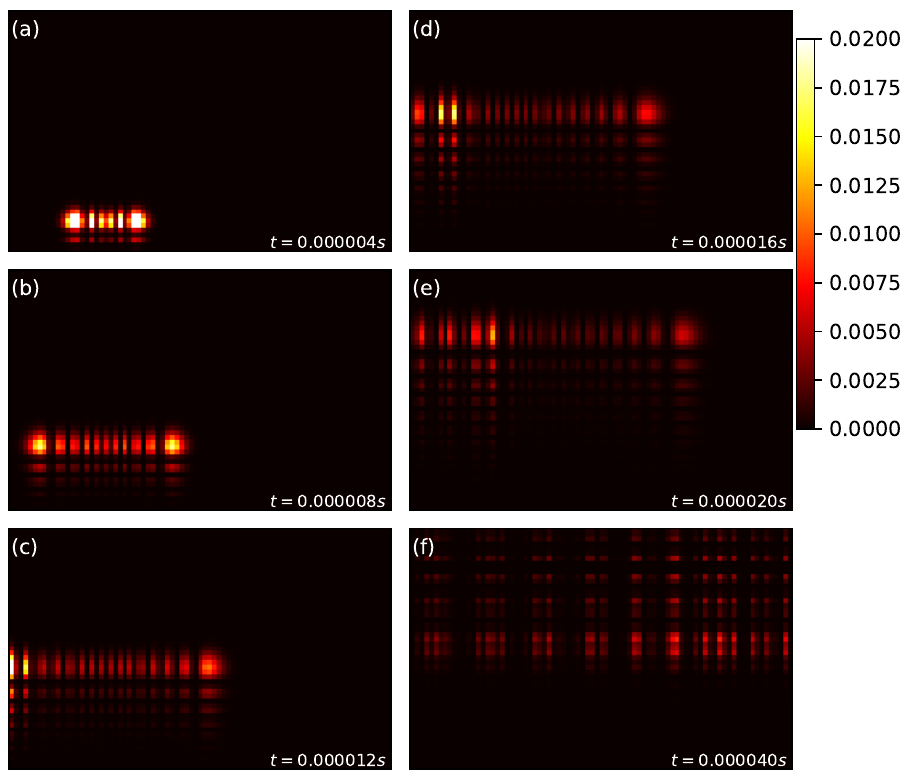} 
		\caption{(online color) Probability distributions in the case of an asymmetric initial state.}
		\label{fig:PhotonMotionAsymm}
	\end{figure}

We can then more intuitively see the motion of photon on a two-dimensional plane from a two-dimensional density distribution. In Fig. \ref{fig:PhotonMotionSymm}, the probability of the appearance of photon extends from the middle to both sides and from bottom to top. When it meets the border, it bounces towards the middle and down. When propagating from the middle to both sides, the probability peak is always at the two end points, and after a rebound, the probability peak moves closer to the middle. In Fig. \ref{fig:PhotonMotionAsymm}, probability peaks are initially located at both ends, and then the peak on the left first bounces to the right after reaching the boundary first, while the peak on the right bounces later than the peak on the left.

We use the classical Cannon's algorithm to implement distributed computing of two stages (Taylor approximation and unitary evolution) of our computational task, described in detail in Sec. \ref{subsec:CannonAlgorithm} and in Fig. \ref{fig:cannon_algroithm}. The first distributed computing strategy is to split a large matrix into $3\times3=9$ small blocks: each small block is loaded onto the corresponding computing core. The second strategy is to split the large matrix into $9\times9=81$ small blocks. A comparison of the speedup of various distributed computing strategies is shown in the Tab. \ref{tab:ComparisonSpeedup}. We found that when using 9 compute cores and 81 compute cores for parallel computing, the time required to complete a task is less than when no parallel computing strategy is used. However, the speedup when using 81 cores is less than when using 9. This is because as the number of cores increases, the time spent transferring data between cores increases.

	\begin{table*}
		\centering
		\begin{tabular}{|c||c|c||c|c|}
			\hline
			\diagbox{Strategy}{Stage} & \tabincell{c}{Taylor\\approximation} & Speedup & \tabincell{c}{Unitary\\evolution} & Speedup \\
			\hline
			without MPI & 2m41s & 1 & 23h43m53s & 1 \\
			\hline
			$3\times3$ & 36s & 4.472 & 5h19m10s & 4.461 \\
			\hline
			$9\times9$ & 2m25s & 1.110 & 13h22m14s & 1.775 \\
			\hline
		\end{tabular}
		\caption{Comparing the speedup of different distributed computing strategies. The symbols "h", "m" and "s" represent "hour(s)", "minute(s)" and "second(s)" respectively.}
		\label{tab:ComparisonSpeedup}
	\end{table*}

	\section{Conclusion}
	\label{sec:Conclusion}
	
	Through supercomputing modeling of QED applications, we can find that using distributed computing on supercomputer platforms can effectively reduce memory consumption and time costs, which will help us solve the curse of dimensionality problem caused by complex quantum systems. The curse of dimensionality, raised by R.E. Bellman \cite{Bellman1957, Bellman1961}, refers to various phenomena that arise when analyzing and organizing data in high-dimensional spaces that do not occur in low-dimensional settings such as the three-dimensional physical space of everyday experience. Distributed computing can solve some of the computing problems caused by the curse of dimensionality, and combining approximation methods (such as state space selection method, iterative state vector method \cite{You2023}, etc.) on the basis of distributed computing can significantly reduce computational complexity.
	
	\begin{acknowledgments}
	The presented research was funded by the China Scholarship Council, Numbers: 202108090327, 202108090483. The authors acknowledge the center for collective use of ultra-high-performance computing resources (https://www.parallel.ru/) at Lomonosov Moscow State University for providing supercomputer resources that contributed to the research results presented in this paper.
	\end{acknowledgments}

	\bibliography{bibliography}

\end{document}